\documentclass[twocolumn,showpacs,showkeys,preprintnumbers,amsmath,amssymb]{revtex4}

\usepackage{graphicx}
\usepackage{dcolumn}
\usepackage{bm}

\begin{document}

\preprint{APS/123-QED}

\title{Compensation temperature of 3d mixed ferro-ferrimagnetic ternary alloy}

\author{Ebru K{\i}\c{s}-\c{C}am$^1$}
\author{Ekrem Aydiner$^2$}%
\email{ekrem.aydiner@deu.edu.tr}
\affiliation{%
$^1$Department of Physics, Dokuz Eylul University, 35160 \`{I}zmir, Turkey\\
$^2$Department of Physics, Istanbul University, 34134 Istanbul,
Turkey }

\date{\today}

\begin{abstract}
In this study, we have considered the three dimensional mixed
ferro-ferrimagnetic ternary alloy model of the type
AB$_{p}$C$_{1-p}$ where the A and X (X=B or C) ions are alternately
connected and have different Ising spins S$^{A}$=3/2, S$^{B}$=1 and
S$^{C}$=5/2, respectively. We have investigated the dependence of
the critical and compensation temperatures of the model  on
concentration and interaction parameters by using MC simulation
method. We have shown that the behavior of the critical temperature
and the existence of compensation points strongly depend on
interaction and concentration parameters. In particular, we have
found that the critical temperature of the model is independent on
concentration of different types of spins at a special interaction
value and the model has one or two compensation temperature points
in a certain range of values of the concentration of the different
spins.
\end{abstract}

\pacs{75.50.Gg; 75.10.Hk; 75.30.Kz; 05.10.Ln}
\keywords{Compensation
temperature; ferro-ferrimagnetic ternary alloys; Monte Carlo
simulation.}

\maketitle


\section{Introduction}

Molecular-based magnetic materials have recently attracted
considerable interest and study of the magnetic properties
\cite{Liu,He,Gmitra,Ohkoshi1,Ohkoshi2,Ohkoshi3,Ohkoshi4,Sato,Pejakovic,Ohkoshi6,Bobak2,Bobak1,Dely,Ohkoshi5,Buendia,Dely2,Carling,Bobak3,Zhoug}.
A special class of the these materials, the so-called Prussian blue
analogs, such as
(X$_{p}^{II}$Mn$_{1-p}^{II})_{1.5}$[Cr$^{III}$(CN)$_{6}$].nH$_{2}$O
(X$^{II}$=Ni$^{II}$,Fe$^{II}$) \cite{Ohkoshi1,Ohkoshi2} and
(Ni$_{p}^{II}$Mn$_{q}^{II}$Fe$_{r}^{II}$)$_{1.5}$[Cr$^{III}$(CN)$_{6}$].nH$_{2}$O
\cite{Ohkoshi3} which exhibit many unusual properties, for instance,
occurrence of one \cite{Ohkoshi1} or even two \cite{Ohkoshi3}
compensation points, magnetic pole inversion
\cite{Ohkoshi2,Ohkoshi4}, the photoinduced magnetization effect
\cite{Sato,Pejakovic}, inverted magnetic hysteresis \cite{Ohkoshi6}.
These ternary alloys have ferromagnetic-ferrimagnetic properties
since they include mixed both ferromagnetic ($J>0$) and
antiferromagnetic ($J<0$) superexchange interactions between the
nearest-neighbor metal ions. The theoretical investigations of these
systems are difficult because of their structural complexity.
However, to obtain magnetic properties of the molecular-based
magnetic materials, up to now, these systems have been studied by
using effective-field theory \cite{Bobak2}, mean field theory
\cite{Bobak1,Dely,Ohkoshi5} and Monte carlo simulation (MC) methods
\cite{Buendia,Dely2,Carling}.

In this study we consider three dimensional ferro-ferrimagnetic
AB$_{p}$C$_{1-p}$ ternary alloy, consisting of three different Ising
spins A=3/2, B=1, and C=5/2, which corresponds to the Prussian blue
analog of the type
(Ni$_{p}^{II}$Mn$_{1-p}^{II})_{1.5}$[Cr$^{III}$(CN)$_{6}$].nH$_{2}$O
\cite{Ohkoshi1}. In this system, the coupling Cr-Ni is ferromagnetic
and Mn-Cr is antiferromagnetic. Our aim, in this study, is to
clarify the effects of the concentration and the interaction
parameters on the magnetic behavior of the three dimensional ternary
alloy model by using MC simulation method.

\section{The Model and Its Simulation}

Three dimensional ferro-ferrimagnetic AB$_{p}$C$_{1-p}$ Ising model
consists of two interpenetrating cubic sublattices as seen in
Fig.\,1. It can be assumed that the A ions are located on the first
cubic sublattice and the B and C ions are randomly distributed on
the second cubic sublattice with the concentration $p$ and $1-p$,
respectively. Also, to construct a Hamiltonian for this system, the
ion A can be represented by spin S$^{A}$, and on the other hand,
ions B and C can be represented by Ising spins S$^{B}$ and S$^{C}$,
respectively. If the interactions between nearest neighbors can be
chosen such as A ions ferromagnetically interact with B, on the
other hand, antiferromagnetically interact with C ions, thus, spins
of the Prussian blue analog of the type
(Ni$_{p}^{II}$Mn$_{1-p}^{II})_{1.5}$[Cr$^{III}$(CN)$_{6}$].nH$_{2}$O
can be represented by this model where S$^{A}$, S$^{B}$ and S$^{C}$
correspond to Cr, Ni and Mn, respectively. In this study we also
consider next-nearest neighbor interactions between spins S$^{A}$.

The Hamiltonian of the considered system can be written in the form
\begin{eqnarray}\label{1}
H=-\sum_{<nn>}S_{i}^{A}[J_{AB}S_{j}^{B}\varepsilon_{j}+J_{AC}S_{j}^{C}(1-\varepsilon_{j})]
\nonumber\\ -J_{AA}\sum_{<nnn>}S_{i}^{A}S_{k}^{A}
\end{eqnarray}
where S$^{A}=\pm 3/2,\pm 1/2$ for A, S$^{B}=\pm 1,0$ for B and
S$^{C}=\pm 5/2,\pm 3/2, \pm 1/2$ for C, on the other hand,
$\varepsilon_{j}$ is a random variable which takes the value of
unity if there is a spin X (S$^{B}$ or S$^{C}$) at the site $j$, if
it not is zero. In Eq.\,(1), the first sum is over the
nearest-neighbor and the second one is over the next-nearest
neighbor spins. In this Hamiltonian the nearest neighbor
interactions are chosen as $J_{AB}>0$ and $J_{AC}<0$, and the
next-nearest neighbor interactions are chosen as $J_{AA}>0$.

In order to show the effects of the concentration $p$ and the
interaction parameters on the compensation and critical temperature
of the three dimensional ternary alloy model, we simulate the
Hamiltonian given by Eq.\,(1). To simulate this model, we employed
Metropolis Monte Carlo simulation algorithm \cite{Binder} to the
$L\times L\times L$ three-dimensional lattice with periodic boundary
conditions for $L=10$, $12$, $16$, $20$, $24$. One of the cubic
sublattice is fully decorated with spin S$^{A}$, and spins S$^{B}$
and S$^{C}$ are randomly distributed on the other cubic sublattice
with the concentration $p$ or $1-p$, respectively. All initial spin
states in the $L\times L\times L$ three-dimensional lattice are
randomly assigned. Configurations are generated by making
single-spin-flip attempts, which were accepted or rejected according
to the Metropolis algorithm. To calculate the averages, data, over
20 different spin configuration, is obtained by using $50000$ Monte
Carlo steps per site after discarding $10000$ steps.

The sublattice average magnetizations per site are obtained by
\begin{subequations}
\begin{equation}
M_{A}=\frac{2}{L^{3}}
\left\langle\sum_{i}^{L^{3}/2}S_{i}^{A}\right\rangle,\qquad
\end{equation}
\begin{equation}
M_{B}=\frac{2}{L^{3}}
\left\langle\sum_{j=1}^{N_{B}}S_{j}^{B}\right\rangle, \qquad
\end{equation}
\begin{equation}
M_{C}=-\frac{2}{L^{3}}
\left\langle\sum_{j=1}^{N_{C}}S_{j}^{C}\right\rangle
\end{equation}
\end{subequations}
where $N_{B}$ denotes the number of B ions $N_{B}=pL^{3}/2$, whilst
$N_{C}$ represents the number of C ions $N_{C}=(1-p)L^{3}/2$ on the
same cubic lattice. Total magnetization per site is given by
\begin{equation}\label{3}
M=\frac{1}{2}\left(  M_{A}+M_{B}+M_{C}\right) \ .
\end{equation}

\section{Results and Discussion}
\begin{figure}
\includegraphics[width=5cm,height=5cm,angle=0]{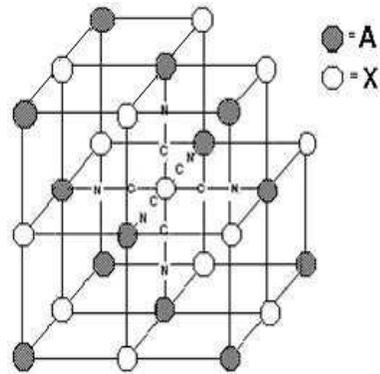}\\
\caption{The crystallographic structure of prussian blue analog with
two interpenetrating cubic lattices.}
\end{figure}

In this section, we have given the simulation results of the ternary
alloy model AB$_{p}$C$_{1-p}$ and we have also discussed the
dependence of the critical and compensation temperature on the
concentration and other interaction parameters in the Hamiltonian.
Simulation results have been obtained for the system with lattice
size $L=10$, $12$, $16$, $20$ and $24$, however, here, we have only
presented the results of the model with lattice size $L=20$. We also
note that the critical temperature of the system for the different
interaction rates and concentrations have been obtained by using of
the method of the finite-size scaling \cite{Binder}.
\begin{figure}
\includegraphics[width=8cm,height=8cm,angle=0]{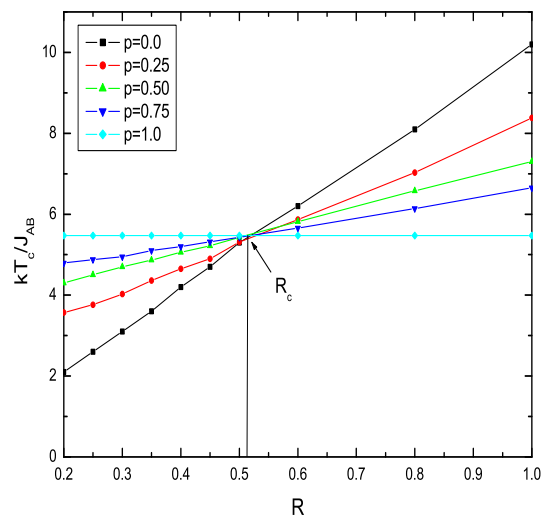}
\caption{Dependence of the critical temperature on interaction ratio
$R$ in the three dimensional ternary alloy AB$_{p}$C$_{1-p}$ for
different values of $p$ when $J_{AA}=0.0$.}
\end{figure}

In a recent study \cite{Buendia} it was reported that two
dimensional ternary alloy model does not show a compensation
temperature point when there is no next-nearest neighbor
interactions term in the Hamiltonian i.e., $J_{AA}=0$. However, our
simulations show that the system has a compensation point for all
$R$ (we set $R=|J_{AC}|/J_{AB}$) values in interval of $0.1\leq R
\leq 2.642$ at $p=0$ when $J_{AA}=0$. This point will be considered
below. Now, in order to compare with the previous results
\cite{Buendia,Bobak3}, in Figs.\,2 and 3 we discuss the dependence
of the critical temperature of the three dimensional ternary alloy
model AB$_{p}$C$_{1-p}$ on interaction rate $R$ and concentration
$p$ for next-nearest neighbor interactions i.e., $J_{AA}=0$.
\begin{figure}
\includegraphics[width=8cm,height=8cm,angle=0]{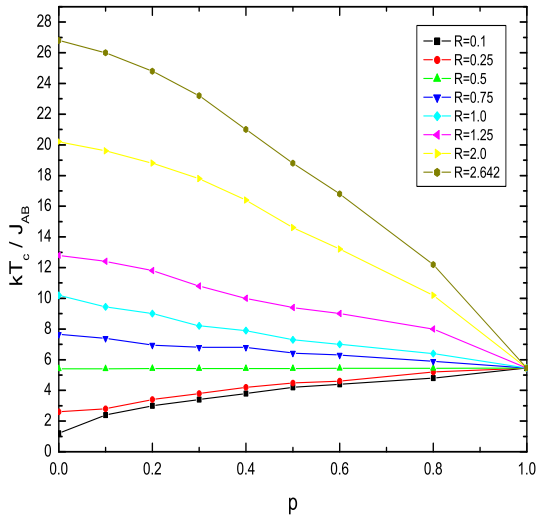}
\caption{Dependence of the critical temperature on the concentration
$p$ in the three dimensional ternary alloy AB$_{p}$C$_{1-p}$ for
several values of interaction ratio $R$ when $J_{AA}=0.0$. The lines
show a part of the second-order transitions separating the
ferrimagnetic and paramagnetic phases.}
\end{figure}

In Fig.\,2, the critical temperature of the three dimensional
ternary alloy model has been plotted as a function of $R$ for
various values of $p$ when $J_{AA}=0$. It can be seen from Fig.\,2
that the critical temperature of the system has a linear dependence
on the interaction ratio $R$ and there is a critical behavior at a
special $R$ value. When $R_{c}=R=0.513$, the critical temperature of
the system has a fixed value of $T_{c}=5.47$ for all $p$ values. At
$R_{c}$, the critical temperature of the system does not change with
concentration $p$. This means that neither the spin-1 ions nor
spin-5/2 ions substitution to system change the critical temperature
of the system at $R_{c}$. This critical behavior has been reported
in theoretical and experimental studies \cite{Buendia,Bobak3,Zhoug}.
The value of the $R_{c}$ for ternary alloy AB$_{p}$C$_{1-p}$ whose
spins consist of S$^{A}=3/2$, S$^{B}=1$ and S$^{C}=5/2$ has been
obtained as $R_{c}=0.4781$ in the study based on mean field
approximation \cite{Bobak3} and as $R_{c}=0.49$ in the Monte Carlo
simulation of two dimensional system \cite{Buendia}. Furthermore,
the experimental measurements indicate that there are Prussian blue
analogs at the $R=0.45$ have a $T_{c}$ which is almost independent
of $p$ \cite{Zhoug}. Fig.\,2 also reveals that concentration $p$
plays an important role for the ternary alloy model
AB$_{p}$C$_{1-p}$ since it determine the kinds of the spins and
interactions in the system. For example, when $p=1$ and $p=0$, the
system AB$_{p}$C$_{1-p}$ fully reduces to the ferromagnetic mixed
spin-3/2 and spin-1 and ferrimagnetic mixed spin-3/2 and spin-5/2
Ising system, respectively. As seen in Fig.\,2, although $T_{c}$ of
the system is independent of $p$ at $R_{c}$, however, the total
magnetization of the system may considerably change owing to
relatively small variation of the concentration $p$. Indeed, for
different values of $p$, the dependence of critical temperature of
the system on the interaction ratio $R$ is very different above and
below of $R_{c}$. This behavior can be explained by the change of
the concentration $p$ in the system. On the other hand, it can be
detected from Fig.\,2 that when $R<R_{c}$, the critical temperature
of the mixed spin-3/2 and spin-5/2 system is smaller than mixed
spin-3/2 and spin-1 system. On the contrary, when $R>R_{c}$, the
critical temperature of the mixed spin-3/2 and spin-5/2 system has
the highest value. On the $T_{c}$ lines, the critical temperature of
the mixed spin-3/2 and spin-1 Ising system is equal to that of the
mixed spin-3/2 and spin-5/2 Ising one.
\begin{figure}
\includegraphics[width=8cm,height=8cm,angle=0]{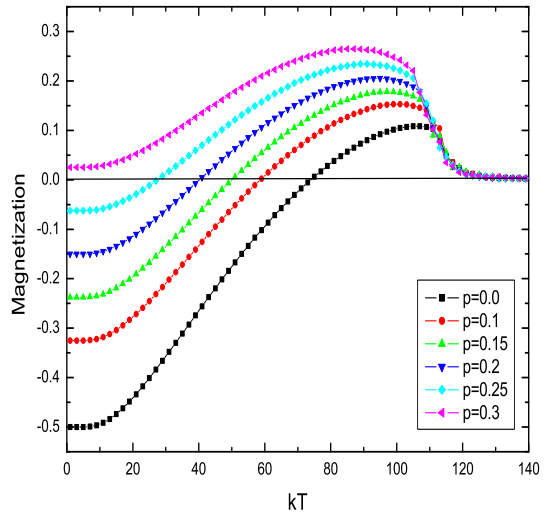}
\caption{Magnetization of the three dimensional ternary alloy
AB$_{p}$C$_{1-p}$ vs temperature for different values of $p$
($J_{AA}=7.5$ and $R=1.0$). }
\end{figure}

In Fig.\,3, the dependence of the critical temperature of the three
dimensional AB$_{p}$C$_{1-p}$ system on the concentration $p$ has
been shown for several values of $R$ when $J_{AA}=0$. The lines
represent part of the second-order phase transition separating the
ferrimagnetic and paramagnetic. Fig.\,3 provides the argument that
the concentration $p$ determines the magnetic features of the system
mentioned above. Indeed, Fig.\,3 clearly shows that the critical
temperature of the system is changed by the concentration $p$ for
fixed values of $R$. As seen from this figure that, when $R<R_{c}$,
the critical temperature of the system linearly increases with
increasing of $p$, whereas, when $R>R_{c}$, the critical temperature
of the system linearly decreases with increasing of $p$ for fixed
values of $R$. However, when the values of $R$ close up $R_{c}$, the
critical temperature of the system more slowly, but linearly, change
with increasing $p$, and at the critical $R_{c}$ value, the critical
temperature of the system denoted by triangle-line in Fig.\,3 is
independent of the concentration $p$. On the other hand, Fig.\,3
also shows that the interaction rate $R$ plays an important role on
the critical temperature of the three dimensional AB$_{p}$C$_{1-p}$
system. Finally we state that the critical temperature of the model
are consistent with previous result \cite{Bobak3}.
\begin{figure}
\includegraphics[width=8cm,height=8cm,angle=0]{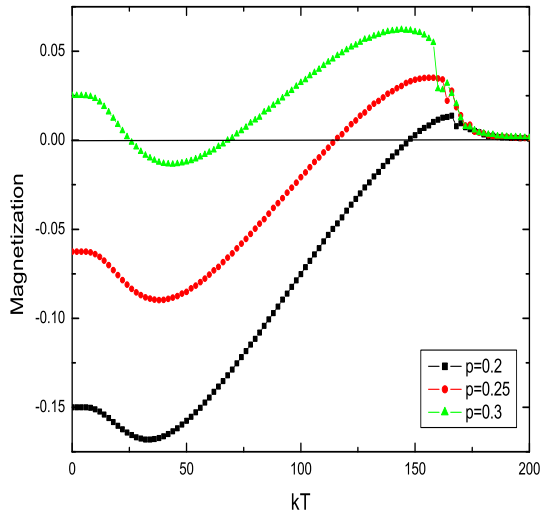}
\caption{Magnetization of the three dimensional ternary alloy
AB$_{p}$C$_{1-p}$ vs temperature for different values of $p$
($J_{AA}=7.5$ and $R=2.642$). }
\end{figure}

In this study we recognize that the three dimensional ternary alloy
model AB$_{p}$C$_{1-p}$ has one compensation behavior for
$J_{AA}=0$, however, for $J_{AA}\neq0$ it has one or multi
compensation points, when other conditions are satisfied. However,
the appearance of the compensation temperature is strongly affected
by the interaction and concentration parameters. Indeed we see in
the present study that the model has not a compensation point for
all values of $p$ and $R$. The dependence of the compensation
temperature behavior on concentration and other interaction
parameters has been discussed below. For discussion, although the
system has been simulated in the intervals of $0.0\leq p <1.0$ and
$0.1\leq R \leq 2.642$, the value of $J_{AA}$ used in the present
study is chosen based on previous theoretical study \cite{Buendia}.
The results of simulation for $R=1.0$ and $R=2.642$ are respectively
represented in Figs.\,4 and 5 for chosen parameters.

One compensation point has been found in the intervals of $0.0\leq p
<0.3$ and $0.1\leq R < 2.642$ when $J_{AA}=7.5$. However, it is seen
that the system has not compensation behavior for the same values of
parameters when $p\geq0.3$. For $R=1.0$ and several values of $p$,
the compensation behavior of the system can be seen from Fig.\,4. On
the other hand, as seen from Fig.\,5, the considered system has a
multi compensation behavior at $R=2.642$ and $p=0.3$ for
$J_{AA}=7.5$ while it has one compensation point for $0.2\leq p <
0.3$. Furthermore, our simulation data introduce that the system
shows compensation behavior at $p=0$ for all values of $R$, when
$J_{AA}=0$. In addition, in the case $J_{AA}=0$, the compensation
point has been found for $R=0.25$ at $p=0.2$, $0.25$; for $R=0.75$
at $p=0.3$; for $R=1.25$ at $p=0.3$; for $R=2.0$ at $p=0.1$, $0.3$,
$0.4$; for $R=2.642$ at $p=0.1$, $0.2$, $0.3$, $0.4$. Whereas, it
has been reported in previous study that there is no compensation
point for $J_{AA}=0$ in two dimensional model \cite{Buendia}.
\begin{figure}
\includegraphics[width=8cm,height=8cm,angle=0]{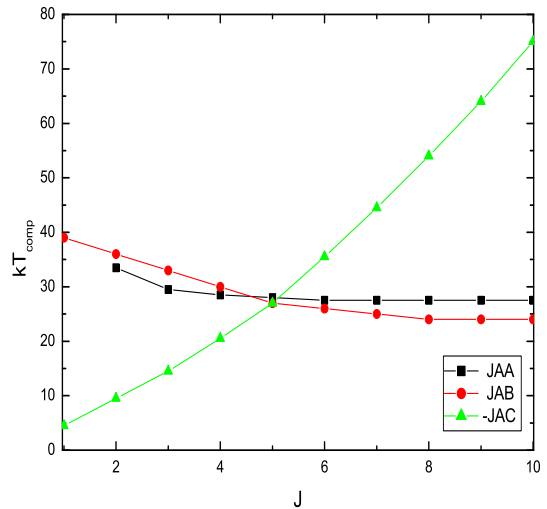}
\caption{Dependence of the compensation temperature $T_{comp}$ on
the interaction parameters in Hamiltonian of ternary alloy
AB$_{p}$C$_{1-p}$ for $p=0.25$.}
\end{figure}


The effect of the interaction parameters on the compensation
behavior of the three dimensional ternary model is also discussed in
Fig.\,6. This figure shows dependence of the compensation
temperature $T_{comp}$ of the model on interaction parameters in
Hamiltonian only for a fixed value of the concentration parameter
$p$ ($p=0.25$). In this figure, square-line represents the behavior
of the compensation point vs $J_{AA}$ for fixed values of
$J_{AB}=5$, $J_{AC}=-5$ and $p=0.25$, circle-line indicates the
behavior of the compensation point vs $J_{AB}$ for fixed values of
$J_{AA}=7.5$ and $J_{AC}=-5$ and $p=0.25$, and on the other hand,
the behavior of the compensation point vs $J_{AC}$ is plotted for
fixed values of $J_{AA}=7.5$, $J_{AB}=5$ and $p=0.25$ with
triangle-line. As seen from Fig.\,6 that for fixed $p$, $J_{AB}=5$
and $J_{AC}=-5$, the compensation temperature decreases slowly as
the strength of the $J_{AA}$ increases. Similarly for fixed $p$,
$J_{AA}=7.5$ and $J_{AC}=-5$, the compensation temperature decreases
slowly with increasing of $J_{AB}$. However, for fixed $p$,
$J_{AA}=7.5$ and $J_{AB}=5$, the compensation temperature
dramatically increases as $|J_{AC}|$ increases. These results
indicate that the compensation temperature has a strong dependence
on the parameter $J_{AC}$ whereas its dependence on $J_{AA}$ and
$J_{AB}$ is relatively weak. The characteristic behavior of the
dependence of the compensation temperature on the parameters of
present model consistent with the results of two dimensional model
\cite{Buendia}.

\section{Conclusion}

In this study, we have considered the three dimensional ternary
model AB$_{p}$C$_{1-p}$ whose spins consist of S$^{A}=3/2$,
S$^{B}=1$ and S$^{C}=5/2$. We have investigated the dependence of
the critical and compensation temperature behavior of the considered
model on concentration and interactions by using MC simulation
method. We have observed that the behavior of the critical
temperature and the existence of compensation points strongly depend
on interaction and concentration parameters. Particularly, we have
found that the critical temperature of the model is independent on
concentration of different types of spins at a critical $R_{c}$
value and the model has one or two compensation temperature points
in a certain range of values of the concentration of the different
spins. We concluded that magnetic properties of the system
AB$_{p}$C$_{1-p}$ can be controlled by changing the relative
concentration of the different species of ions. As a result, we
would like to stress that these theoretical results can be very
useful for designing molecular magnets in experimental studies since
the existence of compensation in the ternary alloy AB$_{p}$C$_{1-p}$
that can be setup by adjusting the proportion of compounds.



\begin{thebibliography}{3}
%
%

\bibitem{Liu} W. M. Liu et al., Phy. Rev. B 65 (2002) 172416.

\bibitem{He} P. B. He and W. M. Liu, Phy. Rev. B 72 (2005) 064410.

\bibitem{Gmitra} M. Gmitra and J. Barnas, Phy. Rev. Lett. 96 (2006) 207205.

\bibitem{Ohkoshi1} S. Ohkoshi, T. Iyoda, A. Fujishima and K. Hashimoto,
Phy.Rev. B 56 (1997) 11642.

\bibitem{Ohkoshi2} S. Ohkoshi, S. Yorozu, O. Sato, T. Iyoda, A. Fujishima
and K. Hashimoto, Appl. Phys. Lett. 70 (1997) 1040.

\bibitem{Ohkoshi3} S. Ohkoshi, Y. Abe, A. Fujishima and K. Hashimoto, Phys.
Rev. Lett. 82 (1999) 1285.

\bibitem{Ohkoshi4} S. Ohkoshi and K. Hashimoto, J. Am. Chem. Soc. 121
(1999) 10591.

\bibitem{Sato} O. Sato, T. Iyoda, A. Fujishima and K. Hashimoto,
Science 271 (1996) 49.

\bibitem{Pejakovic} D. A. Pejakovic, J. L. Manson, J. S. Miller and A. J.
Eipstein, Current Appl. Phys. 1 (2001) 15.

\bibitem{Ohkoshi6} S. Ohkoshi, T. Hozumi and K. Hashimoto,
Phy. Rev. B 64 (2001) 132404.

\bibitem{Bobak2} A. Bob\'{a}k, O. F. Abubrig and D. Horv\'{a}th, Physica
A 312 (2002) 187.

\bibitem{Bobak1} A. Bob\'{a}k and J. Dely, Physica A 341, (2004) 281.

\bibitem{Dely} J. Dely and A. Bob\'{a}k, Physica B 388 (2007) 49.

\bibitem{Ohkoshi5} S. Ohkoshi and K. Hashimoto, Phys. Rev. B 60 (1999)
12820.

\bibitem{Buendia} G. M. Buend\'{\i}a and J. E. Villarroel, J. Magn. Magn. Mater. 310 (2007) 495.

\bibitem{Dely2} J. Dely, A. Bob\'{a}k and M. \v{Z}ukovi\v{c}, Phys. Lett. A 373 (2009) 3197.

\bibitem{Carling} S. G. Carling and P. Day, Polyhedron 20 (2001) 1525.

\bibitem{Bobak3} A. Bob\'{a}k, F. O. Abubrig, T. Balcerzak, Phy.
Rev. B 68 (2003) 224405.

\bibitem{Zhoug} P. Zhoug, D. Xue, H. Lou and X. Chen, Nanoletters 2 (2002) 845.

\bibitem{Binder} K. Binder, in: K. Binder (Ed.), Monte Carlo Methods in Statistical
Physics, Springer, Berlin, 1979.

\end{thebibliography}
\end{document}